\documentclass[11pt,onecolumn]{amsart}
\newtheorem{theorem}{Theorem}[section]
\usepackage[left=1.5cm,top=1.5cm,bottom=1.5cm,right=1.5cm]{geometry}
\usepackage{latexsym}
\usepackage{amssymb,amsmath,amsfonts}
\usepackage[pagewise]{lineno}%\linenumbers
\usepackage{multirow}
\usepackage{graphicx}
\usepackage{cite}
\usepackage[colorlinks = true]{hyperref}
\theoremstyle{definition}
\newtheorem{definition}[theorem]{Definition}

\newtheorem{proposition}[theorem]{Proposition}

\newcommand{\R}{\mathbb R}

\newcommand{\N}{\mathbb N}

\newcommand{\Kcal}{\mathcal{K}}
\newcommand{\Scal}{\mathcal{S}}
\newcommand{\Lcal}{\mathcal{L}}

\newcommand{\Pcal}{\mathcal{P}}
\numberwithin{equation}{section}
 \global\parskip 6pt \oddsidemargin
0.1cm \evensidemargin 0.4cm \headheight 0.1cm \topmargin -2.0cm
\begin{document}
\pagenumbering{arabic}
\title[Fractional Logistic Growth with Memory Effects]{Fractional Logistic Growth with Memory Effects: A Tool for Industry-Oriented Modeling}
\author{M.O. Aibinu$^{1,2,3,*}$, A. Shoukat$^1$, F.M. Mahomed$^3$}
\address{$^{1}$ Department of Mathematics and Statistics, University of Regina, Regina, SK S4S 0A2, Canada}
\address{$^{2}$ Ingram School of Engineering, Texas State University, TX 78666, United States}
\address{$^{3}$ School of Computer Science and Applied Mathematics, University of the Witwatersrand, Johannesburg 2050, South Africa}

\email{$^{*}$moaibinu@yahoo.com (MO Aibinu)}
%\email{$^{*}$moaibinu@yahoo.com | mathew.aibinu@uregina.ca (MO Aibinu)}
%\email{affan.shoukat@uregina.ca (A Shoukat)} 
%\email{fazal.mahomed@wits.ac.za (FM Mahomed)} 
%\dagger


\begin{abstract}
The logistic growth model is a classical framework for describing constrained growth phenomena, widely applied in areas such as population dynamics, epidemiology, and resource management. This study presents a generalized extension using Atangana-Baleanu in Caputo sense (ABC)-type fractional derivatives. Proportional time delay is also included, allowing the model to capture memory-dependent and nonlocal dynamics not addressed in classical formulations. Free parameters provide flexibility for modeling complex growth in industrial, medical, and social systems. The Hybrid Sumudu Variational (HSV) method is employed to efficiently obtain semi-analytical solutions. Results highlight the combined effects of fractional order and delay on system behavior. This approach demonstrates the novelty of integrating ABC-type derivatives, proportional delay, and HSV-based solutions for real-world applications.

\end{abstract}

\maketitle
 %%%%%%%%%%%%%%%%%%%%%%%%%%%%%%%%%%%%%%%%%%%%%%%%%%%%%%%%%%%%%%%%%%%%%%%%%%%%%%%%%%%%%%%%%%%%%%%%%%%%%%%%%%%%%%%%%%%%%%%%%%%%%%%%%%%%%%%%
\section{Introduction}
\par The logistic growth model is a well-known framework for describing how populations grow in limited environments. It explains situations where resources such as space, food, or nutrients place natural limits on expansion. In this model, populations increase quickly at first but eventually slow down and stabilize as they reach the maximum size that the environment can support, called the carrying capacity. This approach allows researchers to study and predict growth patterns in biological, ecological, and social systems, while also providing insight into the key factors that shape population behavior over time. The classical logistic model is characterized by
\begin{equation}\label{ABC8}
	z'(t)=rz(t)\left(1-\frac{z(t)}{\Kcal}\right), \ z(0)=z_0,
\end{equation}
where $z(t)$ represents the variable of interest that grows over time, $r$ is the intrinsic growth rate, and $K$ is the carrying of the environment \cite{Verhulst}. Although the logistic model is widely applied in many fields \cite{HeSYD, KyurkchieN, KucharavyD}, the classical form relies on integer-order derivatives. This structure may not fully capture systems that display memory effects or complex interactions. In many real-world processes, past events influence the present state, creating nonlocal dynamics that require a more refined approach. Previous work addressed this by extending the classical logistic model with integer-order derivatives to a pantograph type, which incorporates delay effects into the model dynamics \cite{AibinuTM}.

Fractional-order models are increasingly used to capture memory effects and nonlocal dynamics in biological, physical, and industrial systems \cite{BaleanuJSM, BaleanuSJ, SaadAB}. However, many earlier studies relied on derivatives with singular kernels, such as the classical Caputo derivative \cite{AibinuME}. These formulations, while mathematically rich, often create difficulties in interpretation and suffer from numerical stiffness caused by the kernel’s singularity \cite{GaoGB}. To address these issues, recent work has introduced fractional operators with nonsingular kernels, including the Caputo-Fabrizio in Caputo sense (CFC) and the Atangana-Baleanu in Caputo sense (ABC) derivatives \cite{NietoJJ, AibinuMFM}. These operators improve realism by modeling systems with smoothly fading memory. CFC-based logistic models have been studied in certain biological and ecological contexts, but the ABC operator remains less explored, especially when combined with delay effects. Unlike the CFC operator, which assumes exponential memory decay, the ABC operator uses a Mittag-Leffler kernel that provides a more flexible memory profile. This makes it particularly suitable for representing complex temporal behaviors such as subdiffusion and long-range correlations.

Furthermore, proportional delays are essential in systems where the response at a given time depends on a scaled version of past states  \cite{PolyaninS, Adomian3}. Such delays arise in biological growth influenced by earlier population densities and in industrial processes governed by lagged control mechanisms. Several analytical and approximation techniques, including Adomian decomposition \cite{Adomian2}, homotopy analysis \cite{Homotopy}, and Laplace-based approaches \cite{Laplace}, have been applied to problems with delays. Among them, the Hybrid Sumudu Variational (HSV) method stands out. This technique combines the Sumudu transform with the variational iteration method to create an efficient iterative approach \cite{LiuC}. Liu and Chen \cite{LiuC} first applied the HSV method to nonlinear equations with integer-order derivatives, while Aibinu and Momoniat \cite{AibinuEM1} extended it to fractional problems with Caputo derivatives, including applications to economic growth models \cite{AibinuM1}.

While delay differential equations have been studied in both classical and fractional frameworks \cite{Rihan, KumarS}, the integration of proportional delay within an ABC-based logistic model introduces an important but largely unaddressed layer of realism. This study develops a novel extension of the classical logistic growth model by combining three key elements: (i) the ABC fractional derivative to capture nonsingular memory effects, (ii) proportional delay to represent time-lagged feedback, and (iii) the HSV method to efficiently resolve the resulting equations. The inclusion of free parameters provides flexibility for modeling diverse growth dynamics in scientific and industrial systems. The HSV method is especially suited for this framework, as it preserves the memory-dependent structure introduced by the ABC derivative while effectively handling the additional complexity from proportional delay. Its superior convergence and reduced computational cost make it a reliable tool for analyzing nonlinear delayed fractional systems. The results highlight how fractional dynamics and delay jointly influence system behavior, demonstrating the combined novelty of ABC-type derivatives, proportional delay, and HSV-based solution strategies. These contributions offer a framework with broad potential applications in fields such as epidemiology, ecological modeling, and industrial process control.

The paper is organized as follows. Section 2 introduces preliminary definitions and summarizes existing results that provide the foundation for this study. Section 3 describes the HSV method used to obtain solutions and also develops the generalized logistic model with ABC fractional derivatives and proportional delay. The analytical results and their discussion are presented alongside the model development, highlighting how memory effects and delay shape system behavior. Finally, Section 4 concludes the paper with a summary of the main findings and their broader implications.

\section{Preliminaries}
Throughout this paper, $\R$ and $\N$ denote the set of real and natural numbers, respectively. 
\begin{definition}
 The Sumudu Transform (ST) is defined over the set
$$\Lambda=\left\{w(t): \exists \ Q, \xi_1, \xi_2 >0, |w(t)|<Qe^{{|t|}/\xi_j}, \mbox{if} \ t\in (-1)^j\times [0, \infty) \right\},$$
by
$$W(u) := \Scal[w(t)] = \int^{\infty}_{0}w(tu)e^{-t}dt, \ u\in (-\xi_1, \xi_2).$$
The ST is related to the well known Laplace transform by the formula
$$W(u) =\frac{1}{u}\int^{\infty}_{0}w(t)e^{-st/u}dt= \frac{1}{u}\Lcal\left(1/u\right),$$
where $\Lcal(u)$ denotes the Laplace transform. ST is linear and has scale and unit preserving properties, which enables it to solve problems without converting to a new frequency domain. This makes it a suitable candidate for integral production-depreciation problems \cite{Belgacem1}.
The transform is well defined for a function that satisfies the Dirichlet conditions and using ST, the Lagrange multiplier can be easily obtained. For any integer order derivative, its ST is expressed as
$$S\left[\frac{dw(t)}{dt}\right]=\frac{1}{u}\left[W(u)-w(0)\right].$$
Generalizing to the $n^{th}$-order derivative, the ST is given as 
$$S\left[\frac{d^kw(t)}{dt^k}\right]=\frac{1}{u^k}\left[W(u)-\displaystyle\sum_{n=0}^{k-1}u^n\frac{d^nw(t)}{dt^n}|_{t=0}\right].$$
\end{definition}

\begin{definition}\label{em20}
	The one parameter Mittag-Leffler function $E_{\mu}(t)$ is defined as
	$$E_{\mu}(t)=\sum_{n=0}^{\infty}\frac{t^n}{\Gamma(n\mu+1)}, \mu >0.$$
	The following results about Mittag-Leffler functions and ST are well known and are presented here without proof \cite{Nanware}:
	\begin{itemize}
		\item [(i)]$$\Scal\left[E_{\mu}\left(-at^{\mu}\right)\right]=\frac{1}{1+au^{\mu}},$$
		\item [(ii)]$$\Scal\left[1-E_{\mu}\left(-at^{\mu}\right)\right]=\frac{au^{\mu}}{1+au^{\mu}}.$$
	\end{itemize}
\end{definition}

\begin{definition}\label{ss5}
Let $a>0, b>0$ be positive real numbers and $0<\mu<1.$ The Atangana-Baleanu fractional derivative of Caputo sense (ABC) of order $\mu$ is defined as \cite{AtanganaK},
$$^{ABC}_aD^{\mu}w(t)=\frac{B(\mu)}{\left(1-\mu\right)}\int^{t}_{a}w'(\xi)E_{\mu}\left(-\frac{\mu}{1-\mu}(t-\xi)^{\mu}\right)d\xi,$$
where $B(\mu)$ is a normalization function that satisfies $B(0) = 1$.
The ST of ABC of order $\mu$ is given by \cite{Bodkhe}, 
\begin{equation}\label{ss6}
\Scal \left[^{ABC}_0D^{\mu}w(t)\right](u)=\frac{B(\mu)}{1-\mu+\mu u^{\mu}}\left(W(u)-w(0)\right).
\end{equation}
\end{definition}

\begin{definition}
Consider the ABC fractional initial-value problem with proportional delay
$$^{ABC}_0D^{\mu}z(t)=f\left(t, z(t), z(\lambda t)\right), z(0)=z_0, 0<\mu<1, 0\leq\lambda\leq 1,$$
where $f(t,x,y)=rx\left(1-\frac{y}{\Kcal}\right).$ A solution is a continuous function
$$z: [0, T]\rightarrow \R,$$
defined on time interval [0, T] with $T>0,$ that satisfies the equivalent Volterra integral equation 
$$z(t)=z_0+\frac{1-\mu}{B(\mu)}\int^{t}_{0}E_{\mu}\left(-\frac{\mu}{1-\mu}(t-\xi)^{\mu}\right)f\left(\xi, z(\xi), z(\lambda \xi)\right)d\xi.$$
A local existence theorem asserts that if $f$ is continuous in $t$ and Lipschitz continuous in $(x,y),$ then there exists some $T^*>0$ such that the initial value problem admits a unique continuous solution on $[0, T^*].$ If, moreover $z(t)$ remains bounded for all $t\geq 0,$ then the solution extends globally with $T=\infty$ \cite{RajmanePG}.
In addition, the solution enjoys Hyers-Ulam stability if a function $u: [0, T]\rightarrow \R$ approximately satisfies the system in the sense that
$$\left|^{ABC}_0D^{\mu}z(t)-f\left(t, z(t), z(\lambda t)\right)\right|\leq \epsilon, t\in [0, T],$$
for some $\epsilon >0,$ then there exists an exact solution $z$ of the system such that
$$|u(t)-z(t)|\leq C\epsilon, ~t\in [0, T],$$
for a constant $C>0$ independent of of $u$ and $\epsilon$ \cite{SousadR}. This means the system is stable under small perturbations of its dynamics, reinforcing both the robustness and physical relevance of the model.
\end{definition}

\begin{proposition}
Let $f, g :[0, \infty)\rightarrow \R,$ then the classical convolution product is given by
$$(f * g)(t)=\int^{t}_{0}f(t-x)g(x)dx.$$
The ST of the convolution product is given by
\begin{eqnarray*}
\Scal \left[(f * g)(t)\right]&=u\Scal[f(t)]\Scal[g(t)]\\
&=uF(u)G(u).
\end{eqnarray*}
\end{proposition}

\begin{definition}
A power series solution is a method for solving differential equations by expressing the solution as an infinite sum. The approach is especially useful when a differential equation defies the standard elementary methods of solutions. Consider a differential equation with a nonlinear term $N[x].$ Writing the solution as a power series,  
$$x(t)=\sum_{n=0}^{\infty}x_nt^n,$$ 
the nonlinear term can be expressed as 
$$N[x(t)]=\sum_{n=0}^{\infty}\Pcal_nt^n,$$
where $\Pcal_n$ are Adomian polynomials \cite{Adomian1,Adomian2}. These polynomials are defined by
$$\Pcal_n=\frac{1}{n!}\left[\frac{d^n}{d{t}^n}f\left(\displaystyle\sum_{i=0}^{\infty}{t}^ix_i\right)\right]\bigg|_{t=0},$$
with the few terms generated as
$$\begin{cases}
\Pcal_0=f(x_0),\\
\Pcal_1=x_1f'(x_0),\\
\Pcal_2=x_2f'(x_0)+\frac{x_1^2}{2!}f''(x_0)\\
\Pcal_3=x_3f'(x_0)+x_1x_2f''(x_0)+\frac{x_1^3}{3!}f'''(x_0)\\
\vdots
\end{cases}$$
Observe that the polynomials $\Pcal_n$ are generated for each nonlinearity such that $\Pcal_0$ depends only on $x_0$, $\Pcal_1$ depends only on $x_0$, and $x_1$, $\Pcal_2$ depends on $x_0,$ $x_1,$ $x_2$ etc. These polynomials facilitate the separation of a nonlinear problem into linear components, allowing for an iterative solution approach.
\end{definition}

\section{Main Results}
In this section, we present a hybrid method of the ST method for a nonlinear problem with ABC and propose the analogs of the classical logistic model, which are given by the ABC and involve time delay.
\subsection{Hybrid of Sumudu transform for fractional derivatives}\label{ss11}
In this study, we apply the HSV method to solve the logistic equation of the ABC type, i.e.,  a nonlocal fractional derivative with a nonsingular kernel. Consider
  \begin{equation}\label{sumud21}
^{ABC}_0D^{\mu}w(t)+\Phi \left[w(t)\right]+\Psi\left[w (t)\right]=g(t),
 \end{equation}
 where $\Phi$ is a linear operator, $\Psi$ is a nonlinear operator and $g(t)$ is a given continuous function. Taking the ST of (\ref{sumud21}) as
 $$ \Scal \left[^{ABC}_0D^{\mu}w(t)\right]=\Scal \left[g(t)-\Phi\left[w(t)\right]-\Psi\left[w(t)\right]\right],$$
and by (\ref{ss6}), we have
$$\frac{B(\mu)}{1-\mu+\mu u^{\mu}}\left(W(u)-w(0)\right)=\Scal \left[g(t)-\Phi\left[w(t)\right]-\Psi\left[w(t-\tau)\right]\right].$$

Therefore, the variation iteration formula takes the form:
\begin{equation}\label{sumud24}
	W_{n+1}(u) = W_n(u)+\phi(u)\bigg(\frac{B(\mu)}{1-\mu+\mu u^{\mu}}(W_n(u)-w(0)) - \Scal \left[g(t)-\Phi\left[w(t)\right]-\Psi\left[w(t-\tau)\right]\right]\bigg), n \in \N.		
\end{equation}

\begin{figure}
\includegraphics[width=10.0cm ,height=8.0cm]{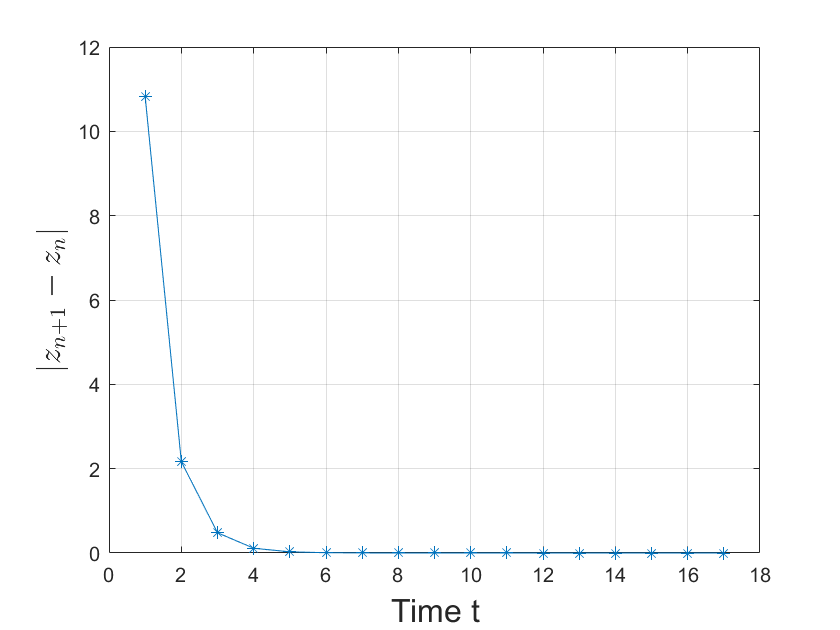}
\caption{Convergence of HSV method.}
\label{ABC19}
\end{figure}

Treating $\Scal \left[\Phi\left[w(t)\right]+\Psi\left[w(t-\tau)\right]\right]$ as the restricted term in taking the classical variation operator on both sides of (\ref{sumud24}) leads to
$$\delta W_{n+1}(u)=\delta W_n(u)+\phi(u)\frac{B(\mu)}{1-\mu+\mu u^{\mu}}\delta W_n(u),$$
which produces the Lagrange multiplier
\begin{equation}\label{emp13}
\phi(u)=-\frac{1-\mu+\mu u^{\mu}}{B(\mu)}.
\end{equation}
Applying (\ref{emp13}) in (\ref{sumud24}) and taking its inverse ST produces
\begin{eqnarray*}
w_{n+1}(t)&=&w_n(t)+{\Scal}^{-1}\bigg[-\frac{1-\mu+\mu u^{\mu}}{B(\mu)}\bigg(\frac{B(\mu)}{1-\mu+\mu u^{\mu}}(W_n(u)-w(0))\\
&&-\Scal \left[g(t)-\Phi\left[w_n(t)\right]-\Psi\left[w_n(t-\tau)\right]\right]\bigg)\bigg]\\
&=&w_1(t)+{\Scal}^{-1}\bigg[\frac{1-\mu+\mu u^{\mu}}{B(\mu)}\Scal \bigg[g(t)-\Phi\left[w_n(t)\right]-\Psi\left[w_n(t-\tau)\right]\bigg]\bigg],
\end{eqnarray*}
with the initial $w_1(t)=w(0).$ Using equation (\ref{ABC8}), Figure \ref{ABC19} illustrates the convergence behavior of the solutions obtained by the HSV method. The figure demonstrates that the HSV method achieves a remarkable rate of convergence, reinforcing its potential as a reliable tool for solving both linear and nonlinear equations. This is particularly valuable in cases where traditional analytical techniques may fail or become intractable. The efficiency and accuracy of the HSV method highlight its suitability for handling fractional models such as the ABC-type logistic equation with proportional delay.

\subsection{Analysis of the Classical Logistic Model}
The classical logistic model (\ref{ABC8}) can be separated to have
\begin{equation}\label{ABC9}
\frac{dz}{z\left(r-\frac{r}{\Kcal}z\right)}=dt.
\end{equation}
Decomposing the left side of (\ref{ABC9})  into partial fractions and integrating leads to
\begin{align*}
\left(\frac{1/r}{z}+\frac{1/\Kcal}{r-\frac{r}{\Kcal}z}\right)dz&= dt\\
\frac{1}{r}\ln |z|- \frac{1}{r}\ln |r-\frac{r}{\Kcal}z|&= t+c_1\\
\ln\left|\frac{z}{r-\frac{r}{\Kcal}z}\right|&= rt + rc_1\\
\frac{z}{r-\frac{r}{\Kcal}z}&= ce^{rt} \text{  where  } c=e^{rc_1}.
\end{align*}

\begin{figure}
\includegraphics[width=10.0cm ,height=8.0cm]{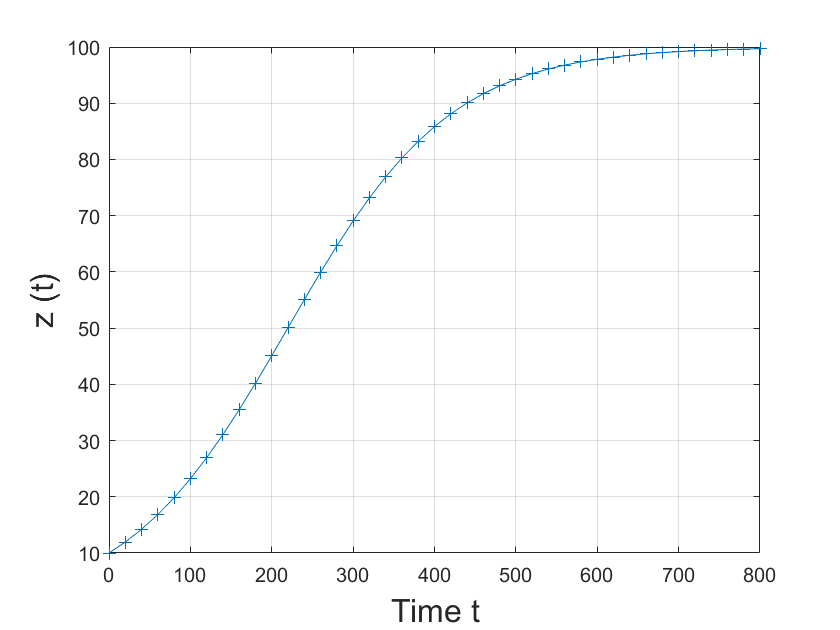}
\caption{Classical logistic model.}
\label{ABC15}
\end{figure}

Simplifying, we have the solution
\begin{eqnarray}\label{ABC10}
z(t)&=&\frac{rce^{rt}}{1+\frac{rc}{\Kcal}e^{rt}}=\frac{rc}{\frac{rc}{\Kcal}+e^{-rt}}.
\end{eqnarray}
Given that $z(0)=z_0, ~z_0\neq \Kcal,$ we have $c = \frac{z_0}{r\left(1-\frac{z_0}{\Kcal}\right)}.$ Substituting for $c$ in (\ref{ABC10}) gives
\begin{equation}\label{ABC11}
z(t) = \frac{z_0}{\frac{z_0}{\Kcal}+\left(1-\frac{z_0}{\Kcal}\right)e^{-rt}}=\frac{z_0\Kcal}{z_0+\left(\Kcal-z_0\right)e^{-rt}}.
\end{equation}
Figure \ref{ABC15} illustrates the behavior of the solution $z(t)$ for the classical logistic model. The population grows rapidly at the early stage when $z(t)$ is much smaller than the carrying capacity 
$\Kcal.$ Over time, as resources become limited, the growth rate slows, and $z(t)$ gradually stabilizes near the carrying capacity, which in this simulation is set to $\Kcal=100.$ This reflects the natural balance imposed by environmental constraints, where competition prevents unbounded growth. From a stability perspective, the classical logistic model (\ref{ABC8}) has two critical points: $z=0,$ which is unstable, and 
$z=\Kcal,$ which is asymptotically stable. This means that any positive initial population will eventually converge to the carrying capacity, capturing the self-regulating nature of the logistic growth process.

\subsection{Generalized Logistic Model with Fractional Derivatives and Proportional Time Delay} Consider a generalized logistic model with proportional time delay and the $ABC$ fractional derivative. This model has the form:
\begin{equation}\label{ABC1}
^{ABC}_0D^{\mu}z(t)=rz(t)\left(1- \frac{z(\lambda t)}{\Kcal}\right), \ z(0)=z_0, \ 0\leq \lambda \leq 1.
 \end{equation} 
We consider two cases of (\ref{ABC1}): the first is when $\lambda = 0$, and the second is when $0<\lambda \leq 1$. For $\lambda=0,$ (\ref{ABC1}) gives
\begin{equation}\label{ABC12}
^{ABC}_0D^{\mu}z(t)=r z(t) \left(1- \frac{z_0}{\Kcal}\right), \ z(0)=z_0.
\end{equation} 
Taking the ST of (\ref{ABC12}) and simplifying, 

\begin{align*}
	\frac{B(\mu)}{1-\mu+\mu u^{\mu}}\left(Z(u)-z_0\right)&=r\left(1- \frac{z_0}{\Kcal}\right)Z(u)\\
	\left[\frac{B(\mu)}{1-\mu+\mu u^{\mu}}-r\left(1- \frac{z_0}{\Kcal}\right)\right]Z(u)&=\frac{B(\mu)z_0}{1-\mu+\mu u^{\mu}}.
\end{align*}
	
\begin{figure}
\includegraphics[width=10.0cm ,height=8.0cm]{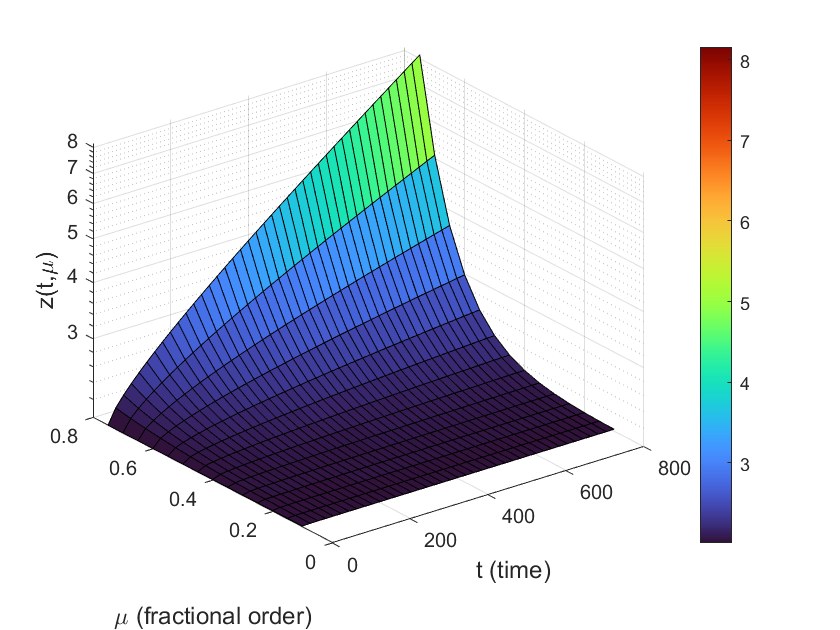}
\caption{Exact solution of the ABC-type logistic model for $\lambda = 0$, showing $z(t,\mu)$ as a function of time $t$ and fractional order $\mu$. Smaller $\mu$ values enhance memory effects, slowing growth, while larger $\mu$ reduce memory influence, leading to faster dynamics approaching the classical logistic behavior. The logarithmic scale highlights gradual acceleration over time.}
\label{ABC16}
\end{figure}

Isolating $Z(u)$, 
\begin{align}\label{ABC13}
Z(u)&= \frac{B(\mu)z_0}{B(\mu)-r\left(1- \frac{z_0}{\Kcal}\right)(1-\mu+\mu u^{\mu})}\nonumber\\
&= \frac{B(\mu)z_0}{B(\mu)+r\left(1- \frac{z_0}{\Kcal}\right)(\mu-1)}\frac{1}{1+\frac{-r\left(1- \frac{z_0}{\Kcal}\right)\mu u^{\mu}}{B(\mu)+r\left(1- \frac{z_0}{\Kcal}\right)(\mu-1)}}.
\end{align}
Taking the inverse ST  of (\ref{ABC13}) and applying Definition \ref{em20} gives
\begin{align}\label{ABC14}
z(t) &= \frac{B(\mu)z_0}{B(\mu)+r\left(1- \frac{z_0}{\Kcal}\right)(\mu-1)}\Scal^{-1}\left[\frac{1}{1+\frac{-r\left(1- \frac{z_0}{\Kcal}\right)\mu u^{\mu}}{B(\mu)+r\left(1- \frac{z_0}{\Kcal}\right)(\mu-1)}}\right]\nonumber\\
	&= \frac{B(\mu)z_0}{B(\mu)+r\left(1- \frac{z_0}{\Kcal}\right)(\mu-1)}E_{\mu}\left(\frac{r\left(1- \frac{z_0}{\Kcal}\right)\mu t^{\mu}}{B(\mu)+r\left(1- \frac{z_0}{\Kcal}\right)(\mu-1)}\right).
\end{align}
which is a fractional and an enhanced form of the Malthusian model \cite{MalthusTR}.
Figure \ref{ABC16} shows the solution of (\ref{ABC12}), which is a fractional enhanced form of the Malthusian growth model \cite{MalthusTR}. The classical Malthusian model faces criticism due to its limitations as an exponential model, which predicts growth if $r > 0$ and decline if $r < 0$. In the context of population growth, this model may hold for a short duration when resources exist in abundance with a low population of species. However, a change in one or more factors, such as weather, food supply, or health can disrupt the growth model. 

Figure \ref{ABC16} shows the exact solution of the ABC-type logistic model when $\lambda=0,$ plotted as a function of time $t$ and fractional order $\mu.$ For smaller values of $\mu,$ the growth is slower, reflecting a stronger memory effect from past states. As $\mu$ increases, the system responds more quickly, and the growth approaches the classical logistic behavior. The logarithmic scale of the population highlights the gradual acceleration of growth over time. Overall, the figure illustrates how the fractional order $\mu$ controls the balance between memory-dependent dynamics and instantaneous growth.

When $0<\lambda \leq 1,$ the model (\ref{ABC1}) is nonlinear. To obtain its solution, we apply the HSV method, which was presented in Section \ref{ss11}. Taking the ST of (\ref{ABC1}) gives

\begin{align}\label{ABC4}
	\frac{B(\mu)}{1-\mu+\mu u^{\mu}}\left(Z(u)-z(0)\right)&=r\Scal\left[z(t)\left(1- \frac{z(\lambda t)}{\Kcal}\right)\right].
\end{align}

Simplifying (\ref{ABC4}) and taking its inverse ST lead to the iteration formula 
\begin{equation*}
	z_{n+1}(t)=z_0 + \frac{1}{B(\mu)}\Scal^{-1}\left[r(1-\mu+\mu u^{\mu})\left(\Scal[z_n]- \frac{1}{\Kcal}\Scal[N[z_n]]\right)\right],
\end{equation*}
where $N[z(t)]=z(t)z(\lambda t)$. Let 
\begin{equation}\label{sse9}
z_n=\sum_{i=0}^nx_i, 
\end{equation}
and the nonlinear term be expressed as
$$N[z]=\sum_{i=0}^{\infty}\Pcal_i, \ \mbox{with} \ \Pcal_i=\frac{1}{i!}\left[\frac{d^i}{d{t}^i}f\left(\displaystyle\sum_{n=0}^{\infty}{t}^nx_n\right)\right]\bigg|_{t=0}.$$

Then the nonlinear term has the Adomian series
$$\begin{cases}
\Pcal_0=f(x_0)=x_0^2, \ \mbox{(since $z(t)=z(\lambda t)$ at $t=0$)}\\
\Pcal_1=x_1f'(x_0)=2x_{0}x_{1},\\
\Pcal_2=x_2f'(x_0)+\frac{x_1^2}{2!}f''(x_0)=2x_{0}x_{2}+x_{1}^{2}\\
\Pcal_3=x_3f'(x_0)+x_1x_2f''(x_0)+\frac{x_1^3}{3!}f'''(x_0)=2x_{0}x_{3}+2x_{1}x_{2}\\
\vdots
\end{cases}$$

%Then the term $x^2$ has the Adomian series
resulting in the iteration formula: 
$$\begin{cases}
x_{n+1}(t)= \frac{1}{B(\mu)}\Scal^{-1}\left[r(1-\mu+\mu u^{\mu})\left(\Scal[x_n]- \frac{1}{\Kcal}\Scal[\Pcal_n]\right)\right],\\
x_0(t)=z(0)=z_0,
\end{cases}$$

Repeated application of this formula produces the sequence:  

$$\begin{cases}
x_0 &=z_0,\\
x_1(t)&= \frac{1}{B(\mu)}\Scal^{-1}\left[r(1-\mu+\mu u^{\mu})\left(\Scal[x_0]- \frac{1}{\Kcal}\Scal[\Pcal_0]\right)\right]\\
  &= \frac{1}{B(\mu)}\left(z_0- \frac{z_0^2}{\Kcal}\right) \Scal^{-1}\left[r(1-\mu+\mu u^{\mu})\right]\\
	&=\frac{rz_0}{B(\mu)}\left(1- \frac{z_0}{\Kcal}\right)\left(1-\mu+ \frac{\mu t^{\mu}}{\Gamma (\mu+1)}\right),\\
	&\\
x_2(t)&= \frac{1}{B(\mu)}\Scal^{-1}\left[r(1-\mu+\mu u^{\mu})\left(\Scal[x_1]- \frac{1}{\Kcal}\Scal[\Pcal_1]\right)\right]\\
	&= \frac{1}{B(\mu)}\Scal^{-1}\left[r(1-\mu+\mu u^{\mu})\left(\Scal[x_1]- \frac{1}{\Kcal}\Scal[2x_0x_1]\right)\right]\\
	&= \frac{1}{B^2(\mu)}\Scal^{-1}\bigg[r(1-\mu+\mu u^{\mu})\bigg(\Scal\left[z_0\left(1- \frac{1}{\Kcal}z_0\right)r\left(1-\mu+ \frac{\mu t^{\mu}}{\Gamma (\mu+1)}\right)\right]\\
	&- \frac{1}{\Kcal}\Scal\left[2z_0^2\left(1- \frac{z_0}{\Kcal}\right)r\left(1-\mu+ \frac{\mu t^{\mu}}{\Gamma (\mu+1)}\right)\right]\bigg)\bigg]\\
	&= \frac{z_0}{B^2(\mu)}\left(1- \frac{z_0}{\Kcal}\right)\left(1-\frac{2z_0}{\Kcal}\right)\Scal^{-1}\bigg[r^2\left((1-\mu)^2+2(1-\mu)\mu u^{\mu}+\mu^2 u^{2\mu}\right)\bigg]\\
	&= z_0\left(\frac{r}{B(\mu)}\right)^2\left(1- \frac{z_0}{\Kcal}\right)\left(1-\frac{2z_0}{\Kcal}\right)\left((1-\mu)^2+2(1-\mu)\frac{\mu t^{\mu}}{\Gamma (\mu+1)} +\frac{\mu^2t^{2\mu}}{\Gamma (2\mu+1)}\right),\\
	&= z_0\left(\frac{r}{B(\mu)}\right)^2\left(1- \frac{z_0}{\Kcal}\right)\left(1-\frac{2z_0}{\Kcal}\right)\left(1-\mu+ \frac{\mu t^{\mu}}{\Gamma (\mu+1)}\right)^2,\\
	&\\
	x_3(t)&= \frac{1}{B(\mu)}\Scal^{-1}\left[r(1-\mu+\mu u^{\mu})\left(\Scal[x_2]- \frac{1}{\Kcal}\Scal[\Pcal_2]\right)\right]\\
	&= \frac{1}{B(\mu)}\Scal^{-1}\left[r(1-\mu+\mu u^{\mu})\left(\Scal[x_2]- \frac{1}{\Kcal}\Scal[2x_{0}x_{2}+x_{1}^{2}]\right)\right]\\
	&= z_0\left(\frac{r}{B(\mu)}\right)^3\left(1- \frac{z_0}{\Kcal}\right)\left(1-\frac{5z_0}{\Kcal}+\frac{5z_0^2}{\Kcal^2}\right)\Scal^{-1}\bigg[\left((1-\mu)^2+2(1-\mu)\mu u^{\mu}+\mu^2 u^{2\mu}\right)(1-\mu+\mu u^{\mu})\bigg]\\
	&= z_0\left(\frac{r}{B(\mu)}\right)^3\left(1- \frac{z_0}{\Kcal}\right)\left(1-\frac{5z_0}{\Kcal}+\frac{5z_0^2}{\Kcal^2}\right)\Scal^{-1}\bigg[(1-\mu)^3+3(1-\mu)^2\mu u^{\mu}+3(1-\mu)\mu^2 u^{2\mu}+\mu^3 u^{3\mu}\bigg]\\
	&= z_0\left(\frac{r}{B(\mu)}\right)^3\left(1- \frac{z_0}{\Kcal}\right)\left(1-\frac{5z_0}{\Kcal}+\frac{5z_0^2}{\Kcal^2}\right)\bigg((1-\mu)^3+3(1-\mu)^2\frac{\mu t^{\mu}}{\Gamma(\mu+1)}+3(1-\mu) \frac{\mu^2t^{2\mu}}{\Gamma(2\mu+1)}+ \frac{\mu^3t^{3\mu}}{\Gamma(3\mu+1)}\bigg)\\
	&= z_0\left(\frac{r}{B(\mu)}\right)^3\left(1- \frac{z_0}{\Kcal}\right)\left(1-\frac{5z_0}{\Kcal}+\frac{5z_0^2}{\Kcal^2}\right)\left(1-\mu+ \frac{\mu t^{\mu}}{\Gamma (\mu+1)}\right)^3,\\
	&\\
\vdots
\end{cases}$$

By (\ref{sse9}), the solution of (\ref{ABC1}) is given by

\begin{align}\label{ABC18}
z(t)&= \displaystyle\lim_{n\rightarrow\infty}z_n =\displaystyle\lim_{n\rightarrow\infty} \sum_{i=0}^nx_i(t) \nonumber \\
&= z_0 + z_0\left(\frac{r}{B(\mu)}\right)\left(1- \frac{z_0}{\Kcal}\right)\left(1-\mu+ \frac{\mu t^{\mu}}{\Gamma (\mu+1)}\right)\nonumber\\
&\qquad + z_0\left(\frac{r}{B(\mu)}\right)^2\left(1- \frac{z_0}{\Kcal}\right)\left(1-\frac{2z_0}{\Kcal}\right)\left(1-\mu+ \frac{\mu t^{\mu}}{\Gamma (\mu+1)}\right)^2\\
&\qquad + z_0\left(\frac{r}{B(\mu)}\right)^3\left(1- \frac{z_0}{\Kcal}\right)\left(1-\frac{5z_0}{\Kcal}+\frac{5z_0^2}{\Kcal^2}\right)\left(1-\mu+ \frac{\mu t^{\mu}}{\Gamma (\mu+1)}\right)^3 + \ldots \nonumber\\
&\approx z_0 \displaystyle\lim_{n\rightarrow\infty} \sum_{i=0}^n\left\{\frac{r}{B(\mu)}\left(1- \frac{z_0}{\Kcal}\right)\left(1-\mu+ \frac{\mu t^{\mu}}{\Gamma (\mu+1)}\right)\right\}^i \ \mbox{as} \ n\rightarrow \infty.\nonumber
\end{align}

\begin{figure}
\includegraphics[width=10.0cm ,height=8.0cm]{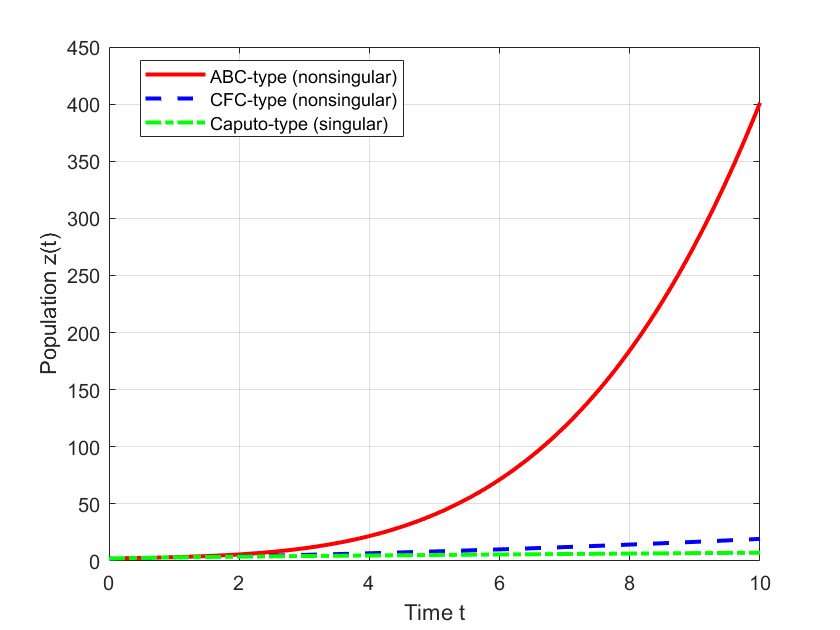}
\caption{Comparison of logistic models with three different fractional operators.}
\label{ABC25}
\end{figure}

Figure \ref{ABC25} compares the logistic growth model with proportional time delay under three different fractional operators. The ABC-type derivative gives a smooth growth curve with gradually fading memory. This behavior comes from the nonsingular Mittag-Leffler kernel, which balances past and present states. The CFC-type model rises faster at the start but shows more rigid dynamics due to its exponential kernel. The classical Caputo derivative, which has a singular kernel, produces sharper fluctuations and less smooth transitions. This highlights its stronger sensitivity to the delay term. The Hyers-Ulam stability analysis shows that small changes in the initial condition or parameters do not greatly affect the ABC-type solution. This confirms the robustness of the model and its reliability for real-world systems. Taken together, the ABC-type operator offers both smoother fading-memory dynamics and stronger stability. These features make it a suitable tool for modeling complex biological, industrial, and social processes with proportional delays.

\begin{figure}
\includegraphics[width=10.0cm ,height=8.0cm]{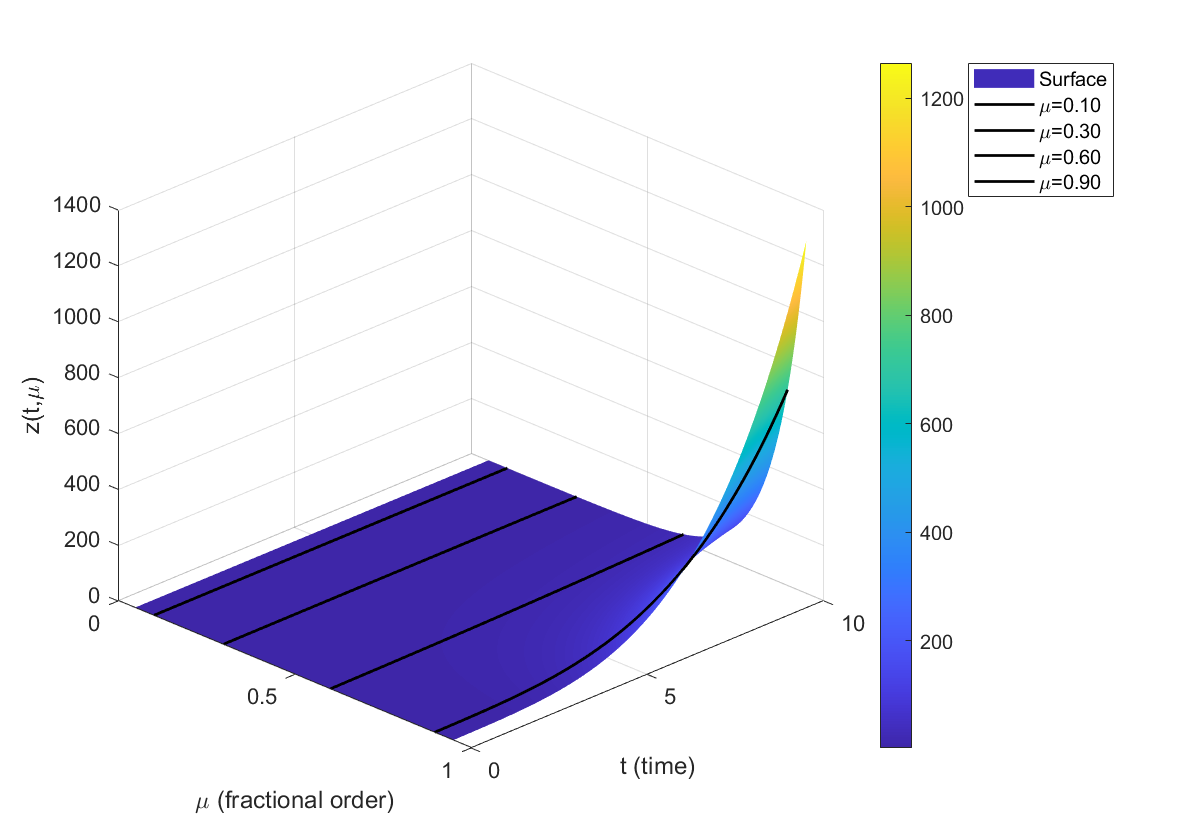}
\caption{3D surface of the ABC-type logistic model with proportional delay, showing $z(t,\mu)$ over time $t$ and fractional order $\mu$. Smaller $\mu$ values enhance memory effects and slow growth, while larger $\mu$ reduce memory, approaching classical logistic dynamics.}
\label{ABC26}
\end{figure}

Figure \ref{ABC26} shows how the fractional order $\mu$ affects the logistic model with the ABC-type derivative and proportional time delay. For small values of $\mu,$ the growth is slower, and the system shows stronger memory of past states. As $\mu$ increases, the effect of memory fades, and the growth curve approaches the classical logistic form. The surface plot highlights this smooth transition from memory-dominated dynamics to more standard growth. This behavior illustrates the flexibility of the ABC-type operator in capturing both long-term memory and near-instantaneous responses, depending on the choice of 
$\mu.$

\begin{figure}
\includegraphics[width=10.0cm ,height=8.0cm]{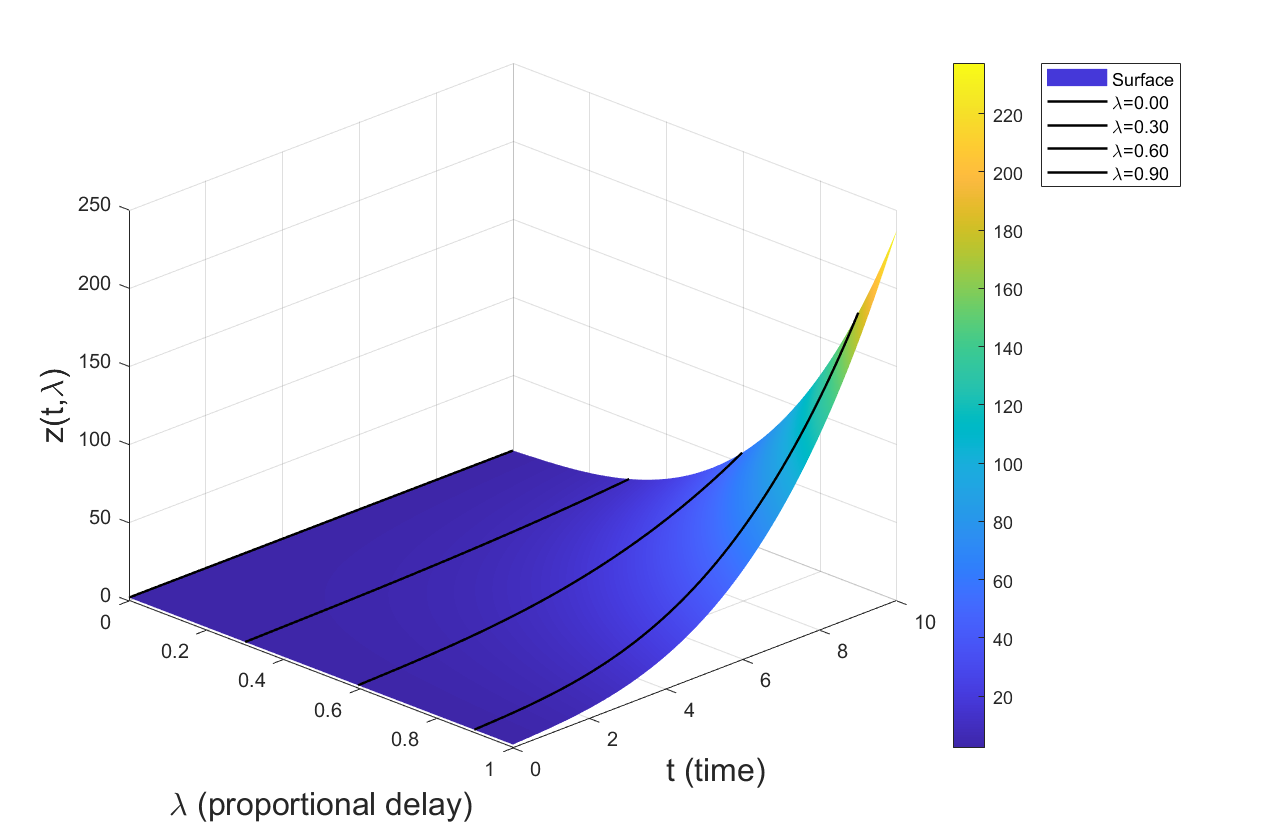}
\caption{3D surface of the ABC-type logistic model showing $z(t,\lambda)$ as a function of time $t$ and proportional delay $\lambda$. Smaller $\lambda$ values enhance the effect of past states, slowing growth, while larger $\lambda$ reduce the delay effect, producing dynamics closer to the classical logistic model.}
\label{ABC27}
\end{figure}

Figure \ref{ABC27} shows how the proportional delay parameter $\lambda$ affects the ABC-type logistic model. For small values of $\lambda,$ the population grows more slowly, indicating a stronger influence of past states. As $\lambda$ increases, the delay effect diminishes, and the growth curve becomes faster and closer to the classical logistic behavior. The surface highlights how varying $\lambda$ allows the model to smoothly transition between memory-dominated dynamics and more immediate responses. This demonstrates the flexibility of the ABC derivative in capturing systems where the timing of past events significantly influences current growth.

\begin{figure}
\includegraphics[width=10.0cm ,height=8.0cm]{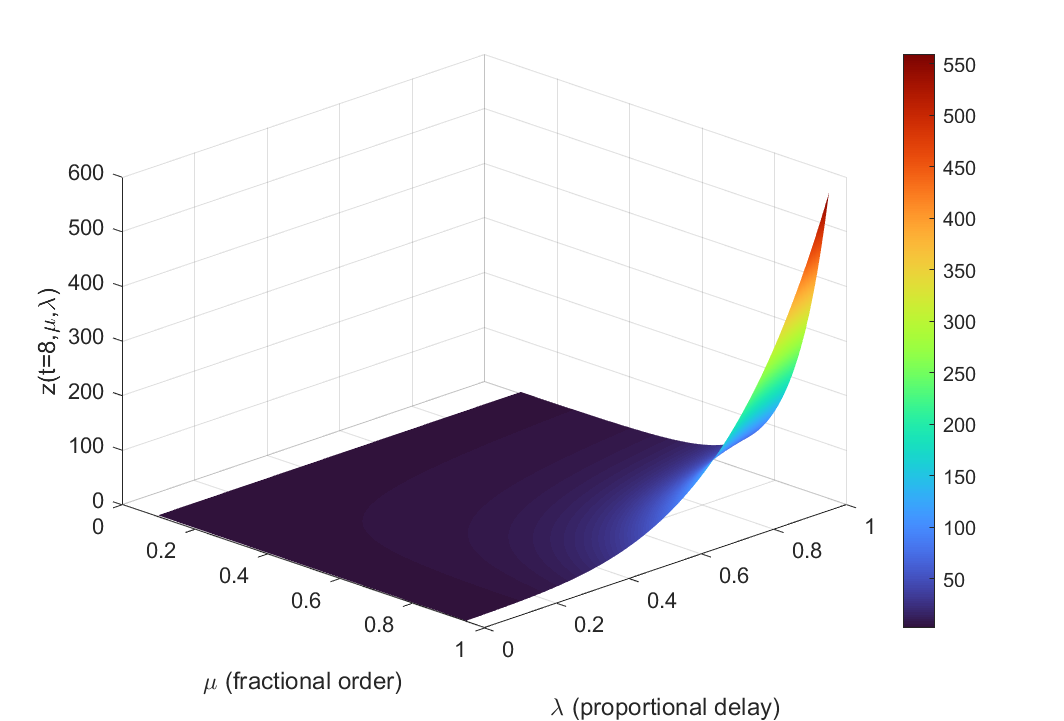}
\caption{3D surface of the ABC-type logistic model showing the combined effects of the fractional order $\mu$ and proportional delay $\lambda$. Smaller $\mu$ values increase memory effects and slow growth, while larger $\mu$ reduce memory influence, approaching classical dynamics. Variation in $\lambda$ further adjusts the growth, highlighting the joint impact of memory and delay on system behavior.}
\label{ABC17}
\end{figure}

Figure \ref{ABC17} presents the graph of (\ref{ABC18}), which is a solution of the generalized logistic model (\ref{ABC1}). The dual aspects of the complexity of (\ref{ABC1}), are the proportional time delay $0<\lambda \leq 1,$ and fractional derivative of the type ABC, where $0.1 \leq \mu \leq 0.9.$ Figure \ref{ABC17} reveals changes in the function $z(t)$ in response to the combination of variation effects in $\mu$ and time $t.$ The 3D surface highlights how both the fractional order $\mu$ and the proportional delay $\lambda$ influence the solution of the ABC-type logistic model at a fixed time. Lower values of $\mu$ increase memory effects, slowing the growth and reducing the solution magnitude. In contrast, larger $\mu$  values reduce memory and allow the system to approach classical logistic behavior. The proportional delay $\lambda$ further modifies this behavior by amplifying or suppressing growth, depending on its strength. Together, $\mu$ and $\lambda$ determine the balance between memory effects and delay, showing that the  dynamics of the model are highly sensitive to both parameters.

\section{Conclusion}
This paper introduces a generalized form of the logistic model that incorporates fractional derivatives of the ABC type together with proportional time delay. By embedding free parameters and arbitrary functions, the model provides a flexible framework for addressing real-world problems, validating experimental data, and testing the robustness of numerical methods. The relevance of this approach spans a wide range of domains where memory effects and nonlocal dynamics are central to system behavior.  

Such phenomena are frequently observed in medicine, for example in modeling tumor growth \cite{BrauerCF, TabassumRB} or the spread of infectious diseases such as COVID-19 \cite{ShenCY, WuDWS}. They also appear in social systems where adaptation to innovation takes place gradually \cite{DovbischukI}, as well as in industrial processes governed by autocatalytic reactions and feedback-based concentration control \cite{XuP}. These areas benefit from models that can accurately capture how past states influence present dynamics.  

The proposed ABC-based logistic model enhances both the theoretical and computational toolkit available for studying systems with memory and delay effects. It advances the mathematical understanding of how memory interacts with delay while also offering a framework for tackling practical challenges in industrial, medical, and social contexts.  

In addition, this study highlights the effectiveness of the HSV method for solving fractional models of this kind. The method shows strong convergence and accuracy, making it a valuable tool for both linear and nonlinear problems. At the same time, it is important to note the boundaries of the approach. The HSV method performs reliably when nonlinear effects are moderate and when the fractional kernel remains smooth. For systems with very stiff dynamics or rapidly changing delay terms, convergence may slow, and refinements such as adaptive iteration or coupling with numerical solvers may be required. These considerations do not lessen the value of the HSV method but instead provide a clearer understanding of the conditions where it is most effective.  

Future work may include testing the model on real industrial or biological data to assess its predictive power, as well as exploring stochastic or network-based extensions. Such directions would further connect the theoretical framework with real-world applications, strengthening the role of fractional modeling in capturing complex memory-driven processes.

\footnotesize
 \vspace{1.0cm}
\noindent {\bf Competing Interests}:\\
The authors declare that they have no relevant financial or non-financial competing interests.\\

\noindent {\bf Abbreviations}:\\
ABC: Atangana-Baleanu fractional derivative of Caputo sense\\
CFC: Caputo-Fabrizio fractional derivative of Caputo sense\\
HSV: Hybrid Sumudu Variational\\
ST: Sumudu Transform\\

\noindent {\bf Disclosure statement}:\\
The authors state that there are no potential conflicts of interest.\\

\noindent {\bf Funding}:\\
There are no funding agencies.

\noindent {\bf Authors’ contributions}:\\
MOA was responsible for the conceptualization of the study. MOA and AS contributed substantially to the manuscript drafting. FMM was responsible for proofreading the manuscript. All authors contributed to the revision of the manuscript and approved the final version.

\section*{Ethical approval}
Ethical approval was not required, as the study did not involve any human participants, animals, or industry-related data.

\section*{Availability of data and materials}
Data sharing is not applicable as no new data were created or analyzed.

%\newpage

\end{document}